# Far-field and near-field investigation of plasmonic-photonic hybrid laser mode

Taiping Zhang, Ali Belarouci*, Ségolène Callard*, Cecile jamois, Xavier Letartre, Celine Chevalier, Pedro Rojo-Romeo, Brice Devif and Pierre Viktorovitch*

*ali.belarouci@ec-lyon.fr;
 segolene.callard@ec-lyon.fr;
 pierre.viktorovitch@ec-lyon.fr

**In last decade, there was a impress development in the fundamental and application researches of localized surface plasmon based structures and devices [1]. Their ultra small mode volume and optical field enhancement ability enable them to find wide applications in the domains of nanophotonics [2], photovoltaics [3], high resolution microscopes [4], surface-enhanced spectrums [5], chemical and biological sensors [6]. However, the extremely tiny sizes make them be hard addressed. Researchers have to face the large energy losses during the exciting processes on the plasmonic devices. Therefore, a key issue is to excite those nano-scale systems more efficient. Herein, we report an approach to achieve this goal via build a plasmonic-dielectric photonic hybrid system. We induce a defect mode based photonic crystal (PC) cavity to work as a intermedium storage as well as a near-field light source to excite a plasmonic nanoantenna (NA). In this way, a plasmonic-photonic nano-laser source is created in present experiment. The coupling condition between the two elements is investigated in far-field and near-field level. We found that the NA reduces the Q-factor of the PC-cavity. Meanwhile, the NA concentrates and enhances the laser emission of the PC-cavity. This novel hybrid dielectric-plasmonic structure may open a new avenue in the generation of nano-light sources, which can be applied in areas such as optical information storage, non-linear optics, optical trapping and detection, integrated optics, etc.**

Localized surface plasmon nanodevices have attracted more and more interest in recent years. Based on their unique property of concentrating light in deep subwavelength space, they may have a lot of potential applications. Most of localized plasmonic structures are metallic nanoparticles. They are usually illuminated by a far-field light source, at best diffraction limited, focus. However, this procedure is not very efficient to address the plasmonic nanodevices. The reason lies in the low coupling rate between the incoming optical beam and the so small metallic nanoparticles in order to make up for the huge optical losses (radiation to free space and metallic absorption). To deal with this challenge, an approach is proposed to induce an intermediate coupling resonator structure between the free space optical beam and plasmonic nanoparticle for providing an appropriate modal conversion of the incoming beam, to optimum address the plasmonic device [7-15]. In our presesnt research, we realize this approach by employ a defect mode photonic crystal cavity that works as the intermediate coupling structure to excite a plasmonic bowtie nanoantenna. We demonstrate a strong coupling between the PC cavity and the NA. In the optimized situations, the NA reduces the Q-factor of the PC-cavity but concentrates the optical energy in near-field level. This research may open a novel efficient route to drive the localize plasmonic devices.

In this research, we approach to achieve a good address of a plasmonic bowtie nanoantenna, which consists of two coupled gold triangles separated by a 20 nm gap, as shown in figure 1a. The bottom edge of the triangle is 140 nm and it is 125 nm high. Geometrical parameters tune an optical response at 1.5 μm. FDTD simulation was made to help us understand the resonance property of the device. We can see the optical field is strongly confined in the gap when the polarization of exciting light is aligned along the gap (figure 1b). Hence we set this direction as the NA axis. Conversely, when the polarization is perpendicular with the gap, the NA cannot concentrate the light into the gap, the optical fields distribute at the four external corners of the NA (figure 1c).

To address the NA, we put the NA on the back bone of a defect mode PC cavity,

which works as the intermedium storage (figure 1d). It is constituted by a triangular array of cylindrical holes with period 420 nm and hole radius 110 nm. The array is patterned on a thin InP slab (thickness 250 nm), which contains four InAsP quantum wells as active material to get laser emission. The slab is supported by a $SiO_2$ substrate. 5 holes of the 2D-PC are omitted to form the defect mode cavity, so called CL5 PC cavity. The fundamental mode of the PC-cavity is around 1.5μm, match with the resonance of the NA. As highlighted in the FDTD simulation (figure 1e to 1g), the optical fields maintained by Ey polarization present in the centre of the cavity and along the long axis of the cavity while the optical fields maintained by Ex polarization distribute along the side of the cavity.

To excite the NA efficiently, we need to achieve a best coupling between the two elements. Consider the polarization sensitivity character of the NA, the position and orientation of the NA should be optimized to coupled with the optical field of PC cavity. We put the NAs at 2 positions on the cavity. The first position is in the centre of the cavity, the orientation of the NA is parallel to the Ey polarization, labeled as NAy (position y, SEM image shown in figure 1h). The second position is near the edge of the cavity and the direction of the NA is parallel to the Ex polarization, labeled as NAx position (SEM image shown in figure 1j). As the FDTD simulation result of the hybrid structures shown in figure 1i and 1k, since the directions of the NAs in both the two cases are parallel to the polarization of optical fields at the regions, the optical fields are localized in the gap of the NAs. And the intensity of the optical fields in the gap of the NA is much stronger than other regions, indicating that the NAs enhances the intensity of the optical fields.

The Far-field optical characterization was performed by micro-photoluminescence spectroscopy at room temperature. The optical setup is described in reference 13. The laser spectra of two hybrid structures (NAy and NAx types) and bare CL5 cavity are shown in figure 2a. The PC cavities of the three structures have the same fabrication parameters. The lasing wavelengths of the three devices are 1514.71 nm (no NA),

1505.47 nm (NAy type) and 1512.52 nm (NAx type) respectively. The variations of the laser peak intensity versus the effective incident pump power of the same structures are shown in figure 2b. It shows that the presence of the NA has an impact on the laser threshold. The laser thresholds of PC cavities with NA are higher than the bare cavity. The laser threshold of the bare PC cavity is about 20 μW. For the hybrid structures, the laser threshold is about 27 μW for the NAy type and 22 μW for the NAx type. These results indicate the influence of the NA on the optical losses (quality factor) of the lasing mode, same as the study on the CL7 hybrid structures in our previous work [13]. This effect can be explained in there is a coupling occurs between the PC and NA modes, the resulting hybrid mode is a linear combination of both modes. Therefore its Q factor lies in between the values of the NA and PC modes. As $Q_{NA} < Q_{PC}$, then $Q_{hybrid} < Q_{PC}$. However, the Q factor decrease may also be due to additional diffraction losses induced by the presence of the NA. In that case, the NA can be seen by the PC mode as a diffusion center. Discriminating between these two explanations cannot be inferred from far field measurements, hence near field experiments are needed to get more information on the spatial properties of the lasing mode.

To investigate the spatial properties of the mode, we employed a homemade Scanning Near-field Optical Microscopy (SNOM) [16]. Another PC cavity without NA was investigated as a reference for evaluating the impact of the NA to the mode distribution. The laser wavelength was measured by SNOM at different locations of the cavity, noted as P1 and P2 (shown in figure 3 a). The wavelengths are $\lambda_{P1}$ = 1514.39 nm and $\lambda_{P2}$ = 1513.62 nm. This wavelength tuning may due to the interaction between the photonic crystal and the near-field probe [17, 18, 19]. And this phenomenon had also been observed in the hybrid structures. The optical field distributions of the laser mode at the two wavelengths measured by SNOM are shown in figures 3c and 3d. This measurement is in agreement with the simulation pattern of the fundamental mode, though the near-field intensity distributions are different and depend on the fixed wavelength. At $\lambda_{P1}$, the region in the centre of the cavity is very bright. At $\lambda_{P2}$, the optical field at the edge of the cavity is very strong.

For the hybrid nano systems, we had done the same characterizations. The SNOM measurement results of a NAy type device and a NAx type device are shown in figure 3e to 3h and figure 3i to 3l, respective. The wavelengths measured at P1 and P2 on NAy cavity are $\lambda_{P1}$ = 1517.38 nm and $\lambda_{P2}$ = 1515.21 nm, and $\lambda_{P1}$ = 1514.8 nm and $\lambda_{P2}$ = 1514.4 nm for NAx cavity. The topography shows the cavities clearly and we can see the feature of the NAs. For both hybrid nanodevices, the near-field optical maps recorded at $\lambda_{P1}$ and $\lambda_{P2}$ are different from the PC cavity without NA and in agreement with the simulation, which indicates a hybrid mode. At $\lambda_{P1}$, the region in the locations of NAs are bright and there are strong hot spots on the NAs. For the NAy structure, the size of the hot spot is smaller than the size of NA. For the NAx structure, the NA is very bright. The regions around the NAs are dark. At $\lambda_{P2}$, the optical fields at the edge of the cavities arise and the intensity are strong. However, the position of NAs are still brighter than the area around them. Especially the NAx structure, the NA still shows the brightest signal. The results of both hybrid structures confirm that the NAs localize the optical fields of the structures.

To confirm the enhancement ability of NA to the optical field intensity in near-field level, we investigate the cross sections of the optical field intensities across the NA. figure 4a and 4b shows the cross section of optical field maintained by Ey polarization at $\lambda$ = 1517.38 nm for the structure of NAy type. The presence of the NA concentrate the light in a very tiny zone where the light intensity is much higher. It is however, very difficult to estimate the enhancement of the field with respect to the case without NA. For the NAx type structure, figure 4c and 4d shows cross sections of optical field Ex at $\lambda$ = 1514.8 nm. On this figure, we clearly see that the NA signal dominates the light intensity in the cavity. The intensity of optical signal of NA is stronger than the signals of other optical field Ex. The results indicates that the NA enhances the optical signal in near-field level.

In conclusion, we design and realize a novel nano-optical device based on the coupling engineering between a CL5 type photonic crystal (PC) cavity and an optical nanoantenna (NA). Optical properties of the hybrid structure are investigated in both far-field and near-field. Through optimise the position and direction of the NA, a plasmonic-photonic crystal laser is demonstrated. The far-field measurements show that a laser signal of hybrid device can be detected. The presence of NA increases the lasing threshold. The SNOM measurement results show that the NA can modify the mode of the hybrid system and localizes the optical field under it. This novel system may have wide potential of applications in integrated opto-plasmonic devices for quantum information processing, as efficient single photon sources or nanolasers, or as sensing elements for bio-chemical species.

**Reference**

[1] E. Ozbay. *Plasmonics : Merging photonics and electronics at nanoscale dimensions,* Science, 311(5758) : 189-193 (2006).

[2] P. Mühlschlegel, H.-J. Eisler, O.J.F. Martin, B. Hecht and D.W. Pohl. *Resonant optical antennas,* Science, 308(5728) : 1607-1609 (2005)

[3] H. A. Atwater and A. Polman. *Plasmonics for improved photovoltaic devices,* Nature Materials, 9 : 205-213 (2010).

[4] N. Fang, H. Lee, C. Sun and X. Zhang. *Sub-diffraction-limited optical imaging with a silver superlens,* science, 308(5721) : 534-537.

[5] S. Nie and S.R. Emory. *Probing Single Molecules and Single Nanoparticals by Surface-Enhanced Raman Scattering,* Science, 277 : 1102-1106 (1997).

[6] R. Elghanian, J.J Storhoff, R.C. Mucic, R.L Letsinger and C.A. Mirkin. *Selective colorimetric detection of polynucleotides based on the distance-dependent optical properties of gold nanoparticles,* Science, 277(5329) : 1078-1081 (1997).

[7] A. Belarouci, T. Benyattou, X. Letartre, P. Viktorovitch. *3D light harnessing based on couping engineering between 1D-2D Photonic Crystal memberanes and metallic nano-antenna,* Optics Express, 18(S3) : A381-A394 (2010).

[8] I.S. Maksymov and A.E. Miroshnichenko. *Active control over nanofocusing with nanorod plasmonic antannas,* Optics Express, 19(7) : 5888-5894 (2011).

[9] F. De Angelis, M. Patrini, G. Das, I. Maksymov, M. Galli, L. Businaro, L.C. Andreani and E. Di Fabrizio. *A Hybrid Plasmonic-Photonic Nanodevice for Label-Free Detection of A Few Molecules,* Nano Lett., 8 : 2321-2327 (2008).

[10] M. Barth, S. Schietinger, S. Fischer, J. Becker, N. Nüsse, T. Aichele, B. Löchel, C. Sönnichsen and O. Benson. *Nanoassembled Plasmonic-Photonic Hybrid Cavity for*

*Tailored Light-Matter Coupling,* Nano Lett., 10 : 891-895 (2010)

[11] F. De Angelis, G. Das, P. Candeloro, M. Patrini, M. Galli, A. Bek, M. Lazzarino, I. Maksymov, C. Liberale, L.C. Andreani and E. Di Fabrizio. *Nanoscale chemical mapping using three-dimensional adiabatic compression of surface plasmon polaritons,* Nature Nanotechnology, 5 : 67-72 (2010).

[12] J. Do, K. N. Sediq, K. Deasy, D. M. Coles, J. Rodríguez-fernández, J. Feldmann and D. G. Lidzey. *Photonic crystal nanocavities containing plasmonic nanoparticles assembled using a laser-printing technique,* Adv. Optical Mater., 1 : 946-951 (2013).

[13] T. Zhang, S. Callard, C. Jamois, C. Chevalier, D. Feng and A. Belarouci. *Plasmonic-photonic crystal coupled nanolaser,* Nanotechnology, 25(31) : 315201 (2014).

[14] Y. Yi, T. Asano, Y. Tanaka, B.-S. Song and S. Noda. *Investigation of electric/magnetic local interaction between Si photonic-crystal nanocavities and Au meta-atoms,* Optics Letters, 39(19) : 5701-5704 (2014).

[15] A. E. Eter, T. Grosjean, P. Viktorovitch, X. Letartre, T. Benyattou and F. I. Baida. *Huge light-enhancement by coupling a bowtie nano-antenna's plasmonic resonance to a photonic crystal mode,* Optics Express, 22(12) :14464-14472 (2014).

[16] T-P. Vo, A. Rahmani, A. Belarouci, C. Seassal, D. Nedeljkovic and S. Callard. *Near-field and far-field analysis of an azimuthally polarized slow Bloch mode microlaser,* Optics Express 18(26) : 26879-26886 (2010)

[17] B. Cluzel, L. Lalouat, P. Velha, E. Picard, E. Hadji, D. Peyrade, F. de Fornel. *Extraordinary tuning of a nanocavity by a near-field probe,* Photonics and Nanostructures - Fundamentals and Applications, 9 : 269-275 (2011)

[18] G. Le Gac, A. Rahmani, C. Seassal, E. Picard, E. Hadji, S. Callard. *Tuning of an active photonic crystal cavity by an hybrid silica/silicon near-field probe,* Optics Express, 17(24) : 21672-21679 (2009)

[19] S. Mujumdar, A.F. Koenderink, T. Sünner, B.C. Buchler, M. Kamp, A. Forchel and V. Sandoghdar. *Near-field imaging and frequency tuning of a high-Q photonic crystal membrane microcavity,* Optics Express, 15(25) : 17214-17220 (2007)

**Acknowledgments**

We acknowledge the financial support from CNRS and China Scholarship Council, as well as the technical staff of Nanolyon platform for support and fruitful discussions.

**Author contributions**

T. Z. and A. B. designed the hybrid nanodevice. T. Z. fabricated the nanodevices and did the SNOM measurement. A. B. did the 3D FDTD simulations. S. C. directed the

SNOM measurement. C. J. and C. C. directed the e-beam lithography. P. R. R. did the RIE process. B. D. deposited the SiO$_2$ hard mask layer. T. Z., A. B., S. C., C. J., X. L. and P. V. discussed the SNOM measurement results. T. Z. wrote the manuscript and revised the manuscript. A. B., S. C. and P. V. supervised the project.

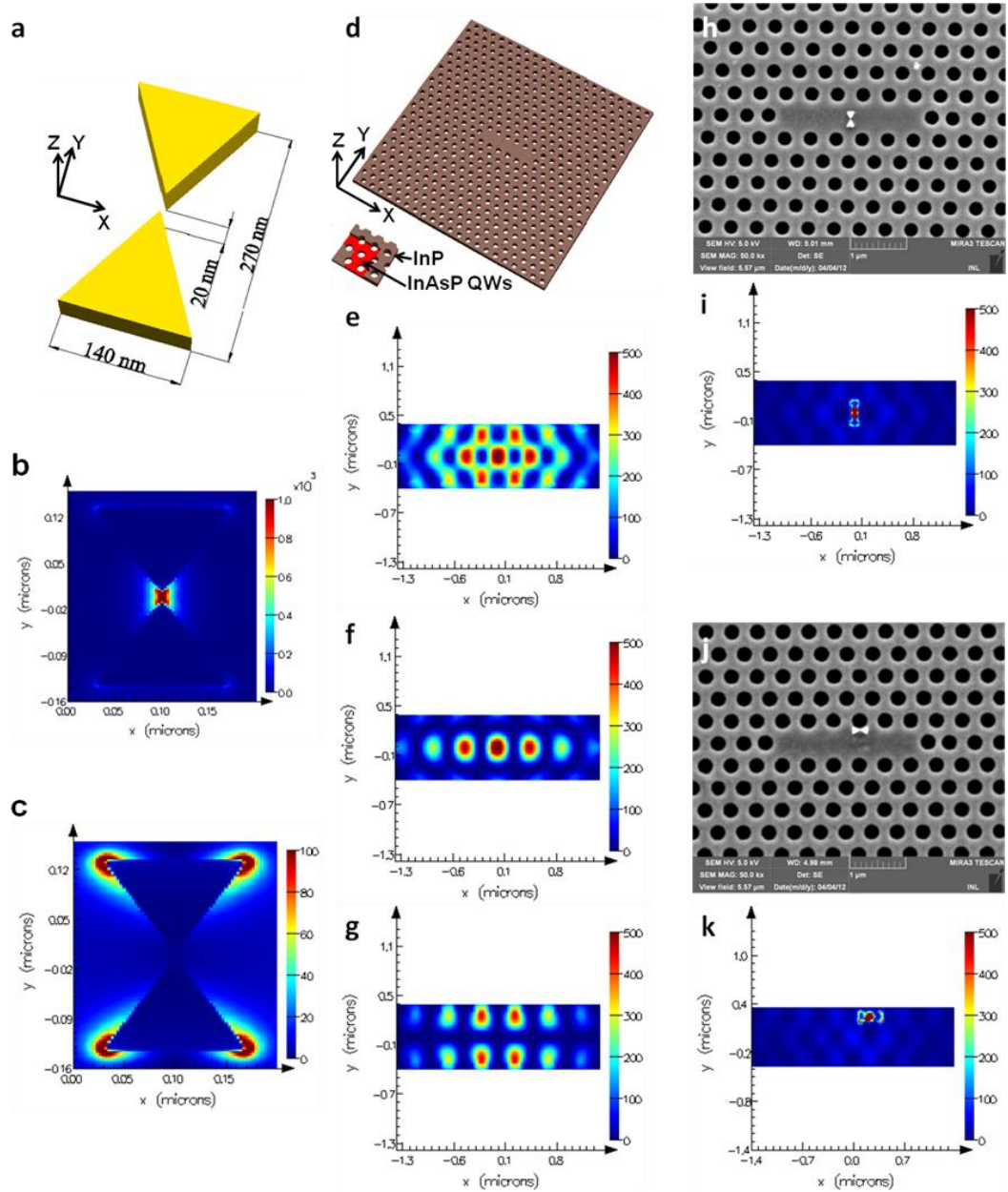

*Figure 1 a) sketch of the bowtie nanoantenna; b) calculated optical field intensity distribution of Ey resonant mode of the nanoantenna; c) calculated optical field intensity distribution of Ey resonant mode of the nanoantenna; d) sketch of CL5 PC cavity; e) spectrum of the laser emission; f) optical field domain of Ey polarization; g) optical field domain of Ex polarization; h) SEM image of hybrid cavity with NA at NAy position; i) Numerical simulation of optical field distribution of CL5 PC cavity with NA at the NAy position; j) SEM image of hybrid cavity with NA at NAx position; k) Numerical simulation of optical field distribution of CL5 PC cavity with NA at the NAx position.*

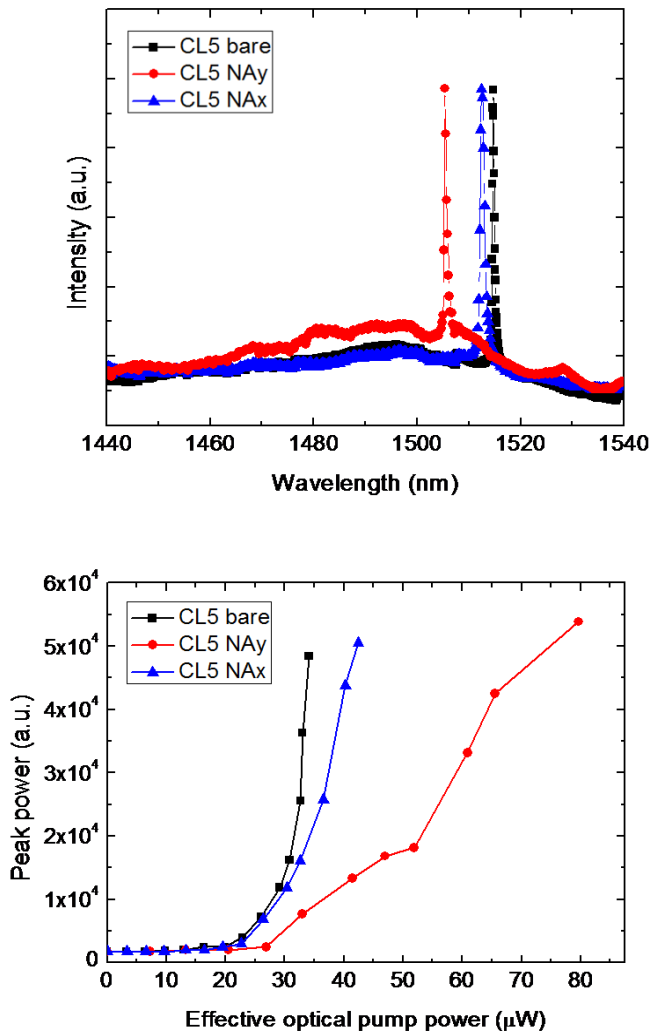

*Figure 2 a) Normalized Spectra of laser emissions of one group of structures. The laser wavelength of bare PC cavity is 1514.71 nm, the laser wavelength of NA at the NAy position is 1505.47 nm, the laser wavelength of NA at the NAx position is 1512.52 nm; b) Threshold of the same group of structures. The threshold of bare PC cavity is 20 µW, the threshold of NA at NAy position is 27 µW, and the threshold of NA at NAx position is 22 µW.*

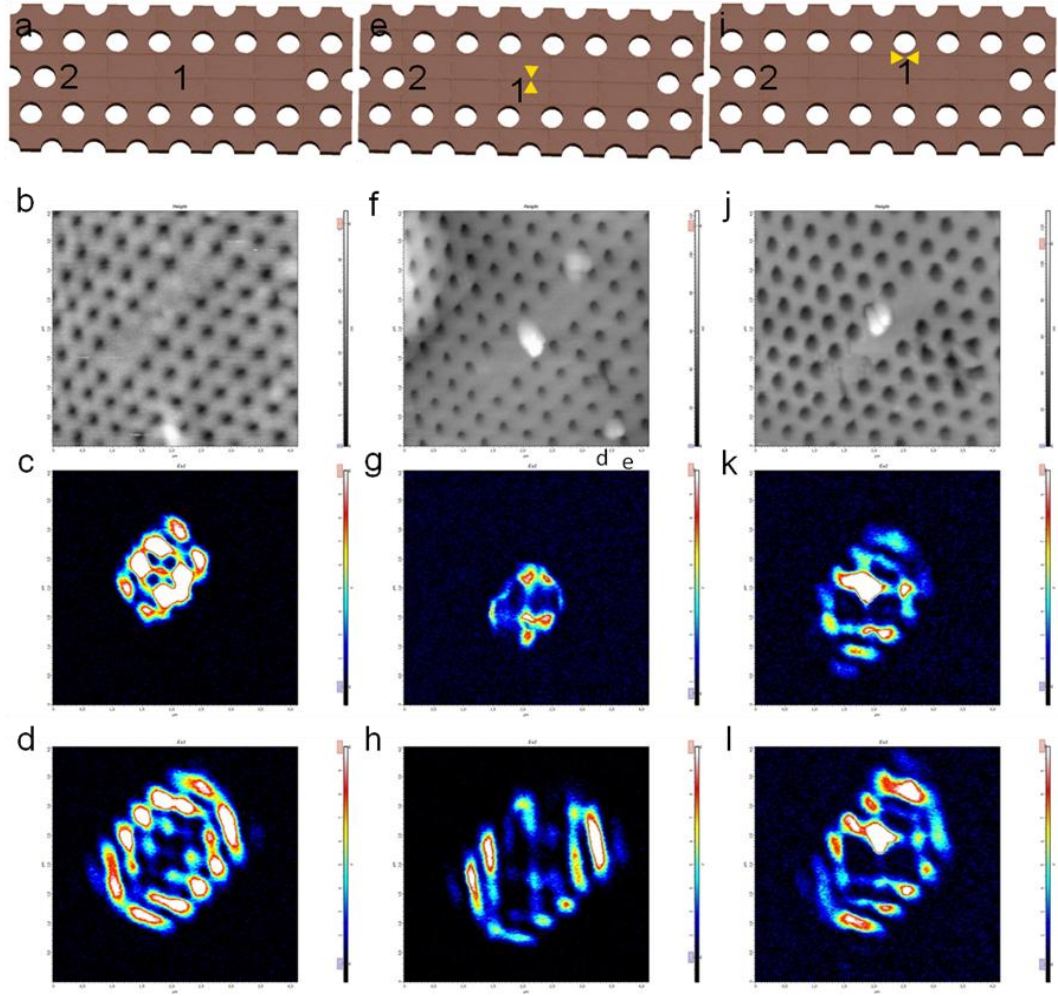

*Figure 3 SNOM measurement result of bare PC cavity, hybrid cavity with NA at the NAy position and NAx position. a), e), i), Different positions of fixed wavelength; b), f), j), topography of the cavity; c), g), k), optical field maps at fixed wavelength of $\lambda_{P1}$; d), h), l), optical field maps at fixed wavelength of $\lambda_{P2}$.*

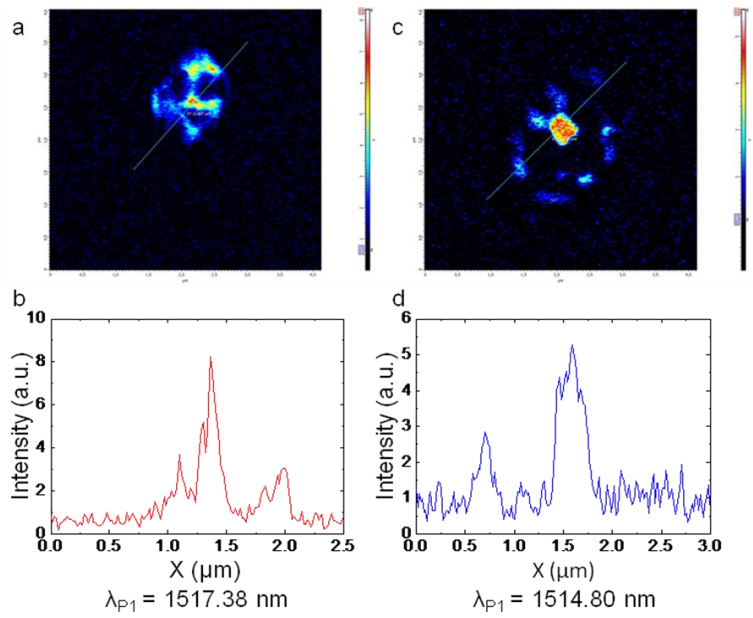

*Figure 4 Cross sections of a), b), hybrid structures with NA at NAy position at λ = 1517.38 nm and c), d), NA at NAx position at λ = 1514.80 nm.*